\documentclass[pra,superscriptaddress,showpacs,papersize=a4paper,floatfix]
{revtex4-2}%
\usepackage[pdftex]{graphicx}
\usepackage{bm}
\usepackage{subcaption}
\usepackage{etoolbox}
\usepackage[usenames,dvipsnames]{color}
\usepackage{mathtools}
\usepackage{amsmath}%
\usepackage{amsfonts}%
\usepackage{amssymb}

\DeclarePairedDelimiterX\braket[2]{\langle}{\rangle}{#1 \delimsize\vert #2}

\newcommand{\bg}{ \begin{gather} }

\newcommand{\eg}{\end{gather}}
\newcommand{\be}{ \begin{equation} }
\newcommand{\ee}{\end{equation}}
\newcommand{\bea}{ \begin{eqnarray} }
\newcommand{\eea}{\end{eqnarray}}

\AtEndEnvironment{thebibliography}{
}

\begin{document}

\title{The butterfly effect in a Sachdev-Ye-Kitaev quantum dot system.}
\author{A. V. Lunkin}
\affiliation{Landau Institute for Theoretical Physics RAS, Moscow 119334, Russia}
\affiliation{HSE University, Moscow 101000, Russia}
\affiliation{ Skolkovo Institute for Science and Technology, Moscow 121205, Russia}

\begin{abstract}
We  study out-of-time-order  correlation function (OTOC) in a lattice extension of Sachdev-Ye-Kitaev (SYK) model with quadratic perturbations.  The results obtained are valid  for arbitrary time scales, both shorter and longer than the Ehrenfest time. We demonstrate that region of well-developed chaos is separated from weakly chaotic region by the "front region", which moves  ballistically  across the lattice. Front velocity is calculated for various system's parameters,
for the first time for SYK-like models.
\end{abstract}
\maketitle

\maketitle

\section{Introduction}
\label{sec: Introduction}
H. Poincaré pointed out the strong dependence of the behaviour of the system on the initial conditions in the study of the  three-body problem. This strange sensitivity was also discovered by  A.Lyapunov.  Edward Lorentz found the same effect during  numerical study of the weather. He proposed to call it "Butterfly Effect". The essence of the effect is following: the "distance" between two close trajectories increase exponentially.   Mathematically, we can write it as:
\begin{eqnarray}
\{q(t),p(0)\}=\frac{\partial q(t)}{\partial q(0)}\sim e^{\lambda_L t}.
\end{eqnarray}

Here $\{\ldots,\ldots\}$-Poisson bracket of the system.  $p$ and $q$ are canonically conjugated momentum and coordinate respectively.  The constant  $\lambda_L$  is called "Lyapunov exponent".  One can generalise this formula for quantum systems. We can consider average of the square of the commutator of the canonical conjugated variables ($p\&q$) to find Lyapunov exponent in such a case:
\begin{eqnarray}
\langle[q(t),p(0)]^2\rangle\sim e^{\lambda_L t}.
\label{eq:Poisson bracket}
\end{eqnarray}
One can expand  brackets in the above formula  (\ref{eq:Poisson bracket}) to obtain  correlation functions with anomalous time-order (no time-ordered nor anti-time-ordered). Larkin and Ovchinnikov were the first who linked such correlators with chaotic properties of the system  \cite{larkin1969quasiclassical}. In general case, we can also consider correlators in the following form:
\begin{eqnarray}
F(t_1,t_2,t_3,t_4)=  \langle X_1(t_1) X_2(t_2) X_3(t_3) X_4(t_4) \rangle,
\label{eq1:OTOC general form}
\end{eqnarray}
here $X_i$-arbitrary  operator. We will assume that $t_1\approx t_3>t_2\approx t_4$.  The behaviour of such correlators   helps to understand chaotic properties  and spreading of the information in quantum systems  \cite{aleiner2016microscopic,sekino2008fast,yoshida2017efficient}.

Let us consider  the possible experimental setup to  measure OTOC. One need to go to the past, perturb the system, wait some time and compare the result with the previous one. Such "experiment" for the classical system was described by R. Bradbury in the sci-fi story "A Sound of Thunder". Modern manageable quantum systems with enormous amount of degrees of freedom  allow to study such correlators.  The one who totally controls evolution of the system can change the sign of the Hamiltonian to  perform evolution back in time.  This experiment was performed by the collaboration Google Quantum Ai  \cite{mi2021information}.

There are a lot of systems where we can write universal expression for  OTOC (\ref{eq1:OTOC general form}) for sufficiently short times.  Namely, we can write one in the form:
\begin{eqnarray}
\frac{F(t_1,t_2,t_3,t_4)}{\langle X_1(t_1)X_3(t_3)\rangle \langle X_2(t_2)X_4(t_4)\rangle } \approx 1- \frac{1}{C}e^{\lambda_L (t_1-t_2)}.
\label{eq: universal behaviour}
\end{eqnarray}
Here $\lambda_L$ is the Lyapunov exponent, as previously, for a quantum system. At the same time $C\gg1$. This formula is applicable for the case when the last term in RHS of (\ref{eq: universal behaviour}) is small to compared to unity; we will refer to this time domain as to the region of weakly developed chaos.  There is the universal bound for Lyapunov exponent for such  systems.  (we assume that $\hbar=1$): 
\begin{eqnarray}
\lambda_L \le 2\pi T.
\end{eqnarray}
This bound was obtained in the work \cite{maldacena2016bound}. The SYK model  is the example of the system which saturates this bound\cite{kitaev2018soft}. We will introduce the Hamiltonian of the SYK model below.  

As was mentioned, above expression (\ref{eq: universal behaviour}) is applicable for short times so it is reasonable to ask: which behaviour does the correlator demonstrate for quite long  times   $t_1-t_2\sim \lambda_L^{-1}\ln C$?  The time scale  $t_E=\lambda_L^{-1}\ln C$ is called the Ehrenfest time. Authors of the work  \cite{maldacena2016conformal} have found the answer for the SYK model. They have shown that after the Ehrenfest time, $t-t_E\gg \lambda_{L}^{-1}$, the correlator $F$ goes to zero exponentially fast.
 Authors of the work  \cite{gu2022two} have considered a behaviour of the OTOC in the general zero-dimension fermionic system. They have considered a  system which obeys to the  universal behaviour (\ref{eq: universal behaviour}) at short times. It was  shown that in this case there is the universal formula for OTOC applicable for arbitrary times: it could be expressed via two fermions correlation function and formula for OTOC for short times.

The typical behavior of the OTOC in the zero-dimensional systems is the following. We can observe the behaviour governed by the formula (\ref{eq: universal behaviour}), for small times, $t\ll t_E$. Chaotic properties are weakly developed in that case. For sufficiently large times, $t-t_E\gg \lambda_{L}^{-1}$, the correlator goes to zero. We will refer to this time domain as to the region of strongly developed chaos.  

Let us consider the system with non-zero  spatial structure.
First of  all, let us clarify which correlator is reasonable to calculate. 
We will consider a case when  operators $X_2$ and $X_4$ act  at point  ${\bf r}_2$;  $X_1$ and $X_3$ act at point  ${\bf r}_1$. We plan to consider OTOC as a function of $\delta t \equiv t_1-t_2$ and $\delta {\bf r} \equiv {\bf r}_1-{\bf r}_2$. The following picture was first described in the paper \cite{aleiner2016microscopic} for usual Fermi liquid.  Assuming time difference $\delta t$ is constant and ${\bf r}_2 ={\bf 0}$, we can highlight two areas in the system.  The first one is the area with weakly developed chaotic properties. This area contains points ${\bf r}_1$ such that  $F=const$.  The second one, where  $F \rightarrow 0$, is the area with strong developed chaotic proprieties.  For example, the points which are far away from the origin belong to the first area. The second area appears for quite long times: $\delta t \ge t_E$. There is the "border"(or "front") between these two areas which moves with increasing time difference. As a result, the second area expands.  The character of this moving is ballistic i.e. the border moves with constant speed (which can depend on the direction of the vector $ \delta {\bf r}$). The same behavior of the front was described in the works \cite{nahum2018operator,von2018operator}.

In the present letter  we investigate the behavior of  OTOC in the  system which  consists of the SYK  "quantum dots".  Nearest SYK quantum dots are coupled by tunneling matrix elements which contribute quadratic perturbation into the total Hamiltonian. We will show that  OTOC in this system obeys the mentioned scenario. In this work, we present the first calculation of the speed of the front in models of such kind. The calculation technique could be applied to other SYK-based models.   

We  note that the authors of the work \cite{gu2017local} considered the behaviour of the OTOC in the SYK-based model.  However, their method is limited by the specific choice of the model. Authors  also did not considered behaviour at times longer then $t_E$.

The structure of the work is the following. We will introduce the model and derive the action in the second section. The third section covers the main properties of the SYK model. We will calculate OTOC and discuss its properties in the fourth section.

\section{Model and its main properties.}
\label{sec: Model and its main properties}

The Hamiltonian of the system has the following form:
\begin{eqnarray}
& H =\sum\limits_{{\textbf r}} \left\{H_{{\textbf r}}+i\sum\limits_{\delta{\textbf r}}  \sum\limits_{i,j}w_{{\textbf r},i;{\textbf r}+\delta{\textbf r},j}  \chi_{{\textbf r},i} \chi_{{\textbf r}+\delta{\textbf r},j}\right\}\qquad 
H_{\textbf r}=\frac{1}{4!} \sum\limits_{i,j,k,l}^N J_{i,j,k,l;{\textbf r}} \chi_{{\textbf r},i} \chi_{{\textbf r},j} \chi_{{\textbf r},k} \chi_{{\textbf r},l}.
\label{eq:Hamiltonian}
\end{eqnarray}
Here 
$\chi_{{\textbf r}}$ is a Majorana fermion with the following commutation relations: $\{\chi_{{\textbf r},i},\chi_{{\textbf r}^\prime,i^\prime}\}=\delta_{i,i^\prime}\delta_{{\textbf r},{\textbf r}^\prime}$. Index $i$ takes values from $1$ to  $N\gg1$. The Hamiltonian (\ref{eq:Hamiltonian}) describes array of "quantum dots". Each  dot is characterized by its position in space, which is defined by the vector ${\textbf r}$. The vector $\delta {\textbf r}$ is a vector between two nearest neighbours. The first term in the Hamiltonian, $H_{\textbf r}$, describes the dynamic inside dots whereas the second term describes the tunnelling of electrons between dots.

The Hamiltonian $H_{\textbf r}$ is the Hamiltonian of the SYK model \cite{kitaev2018soft}.  This Hamiltonian describes a system with $N$ degenerate single-particle states. In this case, the interaction is strong and leads to to the non-Fermi liquid state. The properties of this state were, for example, described in works \cite{banerjee2017solvable,lunkin2021high}.  

Tensors $w$ and $J$ are antisymmetric in indices. Their components are  independent Gaussian variables with zero mean value and dispersion defined as:
\begin{eqnarray}
\langle J_{i,j,k,l;{\textbf r}} ^2\rangle =\frac{3!J^2}{N^3},\qquad  \langle w_{{\textbf r},i;{\textbf r}+\delta{\textbf r},j}  ^2\rangle =\frac{w^2}{N}.
\end{eqnarray}
We assume that the typical energy scale of the interaction inside dots is larger than the typical energy scale of the tunneling i.e.  $J\gg w$

\begin{figure}
	\centering
	\includegraphics[width=0.4\textwidth]{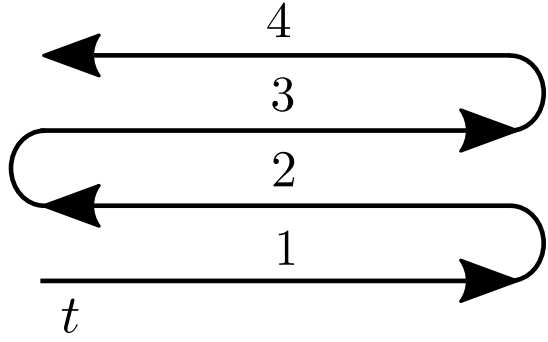}
	\caption{The action is defined as an integral over this contour. One contains two parts that are directed along the time axis and two parts are directed on the opposite side. }
	\label{fig:my_label}
\end{figure}

To calculate  OTOC we will introduce the double Keldysh contour   (Fig. \ref{fig:my_label}). The technical details of this technique   could be found in \cite{aleiner2016microscopic,maldacena2016conformal,keldysh1965diagram}.

 We need to distinguish fields which belong to four different parts of the contour. We will introduce index $\sigma\in \{1,2,3,4\}$ for this purpose. 
For example, we will denote as $\chi_{{\textbf r},i}^{(\sigma)}$ the fermionic field from the $\sigma$th part of the contour .  The correlator we plan to calculate has the following definition:
\begin{eqnarray}
F(t_1,t_2,t_3,t_4)\equiv\frac{1}{N^2}\sum_{i,i^\prime}\langle T_{\mathcal{C}}  \chi_{{\textbf r}_1,i}^{(1)}(t_1) \chi_{{\textbf r}_2,i^\prime}^{(2)}(t_2) \chi_{{\textbf r}_1,i}^{(3)}(t_3)   \chi_{{\textbf r}_2,i^\prime}^{(4)}(t_4)   \rangle 
\label{eq2: OTOC not normalized}
\end{eqnarray}

Here $T_{\mathcal{C}}$ means ordering along the contour. The action on the contour is defined as:

\begin{eqnarray}
S=\sum_{\sigma} \varepsilon_\sigma \int dt \left\{ \sum_{{\textbf r},i}  \frac{1}{2}\chi_{{\textbf r},i}^{(\sigma)}(t) \partial_t \chi_{{\textbf r},i}^{(\sigma)}(t) -H^{(\sigma)}(t)\right\}.
\end{eqnarray}

Here $\varepsilon_1=\varepsilon_3=1=-\varepsilon_2=-\varepsilon_4$ are "signs" of the time direction of different  parts of the contour. After averaging over disorder we have the following action:
\begin{eqnarray}
& S=\sum_{\sigma} \varepsilon_\sigma \int dt_1  dt_0 \delta^\prime(t_1-t_0)  \left\{ \sum_{{\textbf r},i}  \frac{1}{2}\chi_{{\textbf r},i}^{(\sigma)}(t_1)  \chi_{{\textbf r},i}^{(\sigma)}(t_0) \right\}\nonumber \\&+\frac{i}{2}\sum\limits_{{\textbf r}}\sum_{\sigma} \varepsilon_{\sigma_1}  \varepsilon_{\sigma_0} \int dt_1 dt_0 \left\{\frac{J^2}{4 N^3}  \left[\sum_{i}\chi^{(\sigma_1)}_{{\textbf r},i}(t_1)
\chi^{(\sigma_0)}_{{\textbf r},i}(t_0)\right]^4 +\sum\limits_{\delta{\textbf r}}\sum_{i,j}\frac{w^2}{2N} \chi_{{\textbf r},i}^{(\sigma_1)}(t_1) \chi_{{\textbf r},i}^{(\sigma_0)}(t_0) \chi_{{\textbf r}+\delta{\textbf r},j}^{(\sigma_1)}(t_1) \chi_{{\textbf r}+\delta{\textbf r},j}^{(\sigma_0)}(t_0)\right\}. \nonumber
\end{eqnarray} 

Let us define the field  $G$ as:

\begin{eqnarray}
G_{\sigma_1\sigma_0}(t_1,t_0|{\textbf r})=-i\frac{1}{N }\sum_{k} \chi_{{\textbf r},k}^{(\sigma_1)}(t_1) \chi_{{\textbf r},k}^{(\sigma_0)}(t_0).
\label{eq: G definition}
\end{eqnarray}
We also need to introduce a field   $\Sigma$ which is defined as a Lagrangian multiplier to the field $G$. As a result, we can consider the filed $G$ as a new free field in our problem.

We can consider the tunneling term as a perturbation as the tunneling amplitude is much less than the typical energy scale of the interaction inside a dot.  We  use fermions from one dot (\ref{eq: G definition}) to define the field    $G$  as there is no connection between dots in the unperturbed problem. We also introduce the following notation:
\begin{eqnarray}
\Sigma^{(free)}_{\sigma_1\sigma_0}(t_1,t_0)=-\varepsilon_{\sigma_1}\delta_{\sigma_1,\sigma_0} \delta^\prime(t_1-t_0).
\end{eqnarray}
Using new fields and definitions, we can write the action as:

\begin{eqnarray}
& S=-\sum_{{\textbf r},\sigma_1,\sigma_0}  \frac{iN}{2} \int dt_1  dt_0  \Sigma^{(free)}_{\sigma_1\sigma_0}(t_1,t_0) G_{\sigma_1\sigma_0}(t_1,t_0|{\textbf r})\nonumber \\& +\frac{iN}{2}\sum\limits_{{\textbf r}}\sum_{\sigma} \varepsilon_{\sigma_1}  \varepsilon_{\sigma_0} \int dt_1 dt_0 \left\{\frac{J^2 }{4 }  G^4_{\sigma_1\sigma_0}(t_1,t_0|{\textbf r}) -\sum\limits_{\delta{\textbf r}}\frac{w^2}{2} G_{\sigma_1\sigma_0}(t_1,t_0|{\textbf r}) G_{\sigma_1\sigma_0}(t_1,t_0|{\textbf r}+\delta{\textbf r})\right\} +\nonumber\\
&  -\frac{1}{2}\sum\limits_{{\textbf r}}\sum_{\sigma_1,\sigma_0}\int dt_1 dt_0  \Sigma_{\sigma_1\sigma_0}(t_1,t_0|{\textbf r})\left\{ iNG_{\sigma_1\sigma_0}(t_1,t_0|{\textbf r})-\sum_{k} \chi_{{\textbf r},k}^{(\sigma_1)}(t_1) \chi_{{\textbf r},k}^{(\sigma_0)}(t_0)  \right\}. 
\end{eqnarray} 
We need to rewrite expression for OTOC  (\ref{eq2: OTOC not normalized}) using new fields. The correlator  we plan to calculate has the following form in new notations:
\begin{eqnarray}
\Tilde{F}(t_1,t_2,t_3,t_4)= 
\frac{\langle T_{\mathcal{C}} G_{13}(t_1,t_3|{\textbf r}_1) G_{24}(t_2,t_4|{\textbf r}_2)  \rangle }{\langle T_{\mathcal{C}} G_{13}(t_1,t_3|{\textbf r}_1)\rangle  \langle T_{\mathcal{C}} G_{24}(t_2,t_4|{\textbf r}_2)  \rangle}
\end{eqnarray}
The correlators in this form were calculated at papers \cite{maldacena2016conformal,gu2022two} for zero-dimensional models.

 The action we are working with is quadratic in fermionic fields. It means, that we can calculate the functional integral over these fields. It will bring us to the action written in terms of fields $G$ and $\Sigma$.
 
But firstly we would like to introduce a new dimensionless time defined as  $u\equiv 2\pi T t $. Where $T$ is the temperature of our system. We also rescale fields: 
\begin{eqnarray}
\Sigma(t_1,t_0)\equiv J^2 \left(\frac{2\pi T}{J}\right)^{3/2}\Sigma(u_1,u_0) \qquad
G(t_1,t_0)\equiv \left(\frac{2\pi T}{J}\right)^{1/2}G(u_1,u_0)
\end{eqnarray}
We can integrate out fermionic fields and write the action using dimensionless variables as:
\begin{eqnarray}
&\frac{2S}{iN}=-\sum\limits_{{\textbf r}} \sum\limits_{\sigma_1,\sigma_0}  \int du_1  du_0 \biggl\{-\frac{\varepsilon_{\sigma_1}\varepsilon_{\sigma_0} }{4 }  G^4_{\sigma_1\sigma_0}(u_1,u_0|{\textbf r})+\left[\Sigma^{(free)}_{\sigma_1\sigma_0}(u_1,u_0)+\Sigma_{\sigma_1\sigma_0}(u_1,u_0|{\textbf r}) \right] G_{\sigma_1\sigma_0}(u_1,u_0|{\textbf r})  + \nonumber \\
&\left.\sum\limits_{\delta{\textbf r}}\frac{\varepsilon_{\sigma_1}\varepsilon_{\sigma_0}}{2}\left(\frac{w^2}{2\pi T J}\right) G_{\sigma_1\sigma_0}(u_1,u_0|{\textbf r}) G_{\sigma_1\sigma_0}(u_1,u_0|{\textbf r}+\delta{\textbf r})\right\} -tr\ln(\Sigma)
\label{eq3: action Green function}
\end{eqnarray}

The last term in the action  (\ref{eq3: action Green function})  is dominant for sufficiently low temperature $T\ll T_{FL}\sim \frac{w^2}{J}$. The system demonstrates the properties of the Fermi-liquid in this case. We assume that $T\gg T_{FL}$. It means that  the tunneling term could be considered as a perturbation. The action without the tunneling terms is the action of the SYK model. We will consider its properties below.

\section{The main properties of the SYK mode}
\label{sec: The main properties of the SYK mode}

In this Section, we will consider the  behaviour of the fermionic
Green function in the thermal equilibrium within a mean-field approximation. This "saddle-point" solution for the Green  function will  important to find the low-energy action of the SYK model in terms of fluctuations near the saddle-point we are going to find now. These fluctuations  will play the crucial role in the analysis provided by the present Letter.

As we have the large parameter in the problem $N\gg1$ we can consider mean-field equations for fields $G$ and $\Sigma$. We have neglected the tunneling term assuming that the temperature is high enough $T\gg T_{FL}$. The term   $\Sigma^{(free)}$ important for hight temperature $T\sim J$ and for short times $t\sim\frac{1}{J}$.  Assuming that $J\gg T \gg T_{FL}$ we can write the saddle-point equation in the following form:
\begin{eqnarray}
& 
\sum\limits_{\sigma}\int du\Sigma_{\sigma_1\sigma}(u_1,u|{\textbf r})G_{\sigma\sigma_0}(u,u_0|{\textbf r}) = \delta_{\sigma_1,\sigma_0}\delta(u_1-u_0),\nonumber \\
&\Sigma_{\sigma_1\sigma_0}(u_1,u_0|{\textbf r})  = \varepsilon_{\sigma_1}\varepsilon_{\sigma_2}  G^{q-1}_{\sigma_1\sigma_0}(u_1,u_0|{\textbf r}).  
\label{eq: mean-field}
\end{eqnarray}
 Physical value of the parameter q is equal to 4.  It is introduced as continuous parameter  for further convenience.  We would like to write the solution of  above equations in the general form.

These equations have an infinitely big group of symmetries. Let us consider the set of monotonous functions $f_{\sigma,{\textbf r}}(u)$ and a solution  $G_{\sigma_1\sigma_0}(u_1,u_0|{\textbf r})$ to prove it. We can apply the symmetry transformation  defined by the following rule:
\begin{eqnarray}
G_{\sigma_1\sigma_0}(u_1,u_0|{\textbf r})  \mapsto \left[f^\prime_{\sigma_1,{\textbf r}}(u_1)f^\prime_{\sigma_0,{\textbf r}}(u_0)\right]^\Delta G_{\sigma_1\sigma_0}(f_{\sigma_1,{\textbf r}}(u_1),f_{\sigma_0,{\textbf r}}(u_0)|{\textbf r}) 
\label{eq5:symmetry}
\end{eqnarray}

\noindent here $\Delta=\frac{1}{q}$. The new function $G$, obtained by this rule, is also a solution to the mean-field equations. However, we know that the initial Hamiltonian is time-independent. It means that the Green's function depends only on the time difference. 

 Working with a system in the thermal equilibrium we can note that  the  Green's functions on the last two parts of the contour  could be restored from the behaviour of the    Green's functions on the first two. So, in the formula below we assume that   $\sigma\in \{1,2\}$. Taking into account the above comment we can write the translation-invariant solution of the mean-field equations as:
\begin{eqnarray}
& G^{(0)}_{\sigma_1\sigma_0}(u_1,u_0|{\bf r}) = b^\Delta g_{\sigma_1\sigma_0}(u_1-u_0) , \nonumber \\
& \Large g_{\sigma_1\sigma_0}(u>0)\equiv i \left[\frac{s^{\prime}_{(0)}(u_1)s^{\prime}_{(0)}(u_0)}{(s_{(0)}(u_1)-s_{(0)}(u_0))^2}\right]^\Delta \left(\begin{smallmatrix}
-e^{-i\pi \Delta } & e^{i\pi \Delta} \\ -e^{-i\pi \Delta} & e^{i\pi \Delta}
\end{smallmatrix}\right)_{\sigma_1\sigma_0} 
\label{eq6:saddle-point}
\end{eqnarray}
\noindent here $ b = \frac{(1-2\Delta)\tan(2\pi \Delta)}{2\pi }$
and $s_0(u)=e^{u}$, this notation will be useful further.
The solution for   $u<0$ could be restored using symmetry properties of the Green's function $g_{\sigma_1\sigma_0}(u)=-g_{\sigma_0\sigma_1}(-u)$.

The symmetry group of the equations is much bigger than the symmetry group of the solution. The  last one is determined by the following transformations:
\begin{eqnarray}
s_0(u)\mapsto s(u)=\frac{a s_0(u)+b}{c s_0(u)+d}.
\end{eqnarray}

This difference between symmetry groups is typical for a "symmetry breaking" . On the other hand, this symmetry of the mean-field equations  is not exact. Even in the case $w=0$ we have neglected  by the term with $\Sigma^{(free)}$ in the equations. It means that the mentioned symmetry  (\ref{eq5:symmetry}) is so-called asymptotic. It is a typical case for $\sigma$-models (see for example \cite{efetov1999supersymmetry}). We should not take a functional integral over all possible fields $G$ and $\Sigma$. We need to take into account fields that are solutions to  approximate  mean-field equations  (\ref{eq: mean-field}). All these fields could be obtained as a result of the application of the symmetry transformations (\ref{eq5:symmetry}) to the saddle-point solution  ( \cite{efetov1999supersymmetry}).
This reduction is not exact, the fluctuation in the perpendicular direction to this manifold is suppressed by the parameter   $\beta J\gg1$ see  \cite{kitaev2018soft}.

We plan to take functional integral over the  manifold parameterized by the set of functions $f_{\sigma,{\textbf r}} (u)$. The action on this manifold has the form (see \cite{kitaev2018soft}):
\begin{eqnarray}
S=-C_J\sum\limits_{{\textbf{r}},\sigma}\varepsilon_{\sigma}\int du~ Sch\{e^{f_{\sigma,{\textbf r}}},u\}- 
C_w\sum\limits_{\delta{\textbf r},{\textbf r},\sigma_1,\sigma_0}\int du_1du_0\varepsilon_{\sigma_1}\varepsilon_{\sigma_0} g^{(f)}_{\sigma_1\sigma_0}(u_1,u_0|{\textbf r}) g^{(f)}_{\sigma_1\sigma_0}(u_1,u_0|{\textbf r}+\delta{\textbf r}).
\label{eq:action }
\end{eqnarray}
Here  
$C_J=\alpha_S N\frac{2\pi T}{J}$ and  $C_w=\frac{ N w^2}{4\pi J T} $ -  is a heat capacity (up to a factor $2\pi$) of the SYK  model and contribution to the heat capacity from the perturbation respectively, $\alpha_S\approx 0.05$.  The field  $g^{(f)}$ (for $u_1-u_2>0$) is defined as:
\begin{eqnarray}
& g^{(f)}_{\sigma_1,\sigma_2}(u_1,u_0|{\textbf{r}})= \left[f^\prime_{\sigma_1,{\textbf r}}(u_1)f^\prime_{\sigma_0,{\textbf r}}(u_0)\right]^\Delta g_{\sigma_1\sigma_0}(f_{\sigma_1,{\textbf r}}(u_1)-f_{\sigma_0,{\textbf r}}(u_0)) .
\label{eq:fields}
\end{eqnarray}
Here  $Sch\{s(u),u\}$ means Schwartz derivative defined as:
\begin{eqnarray}
Sch\{s(u),u\}\equiv \left(\frac{s^{\prime\prime}}{s^\prime}\right)^\prime - \frac{1}{2}\left(\frac{s^{\prime\prime}}{s^\prime}\right)^2.
\end{eqnarray}
One of the most important properties of this derivative is the following: one can substitute $s$ by a linear fractional transformation of $s$ without changing the result. 

The correlation function we need  to calculate, could be rewritten as follows:
\begin{eqnarray}
\Tilde{F}(u_1,u_2,u_3,u_4)=\frac{\langle T_{\mathcal{C}} g^{(f)}_{1,3}(u_1,u_3|{\textbf{r}_1}) g^{(f)}_{2,4}(u_2,u_4|{\textbf{r}_2})  \rangle}{\langle  g^{(f)}_{1,3}(u_1,u_3|{\textbf{r}_1})\rangle \langle  g^{(f)}_{2,4}(u_2,u_4|{\textbf{r}_2})  \rangle}.
\end{eqnarray}

We plan to use a quadratic action for fluctuations which could be used for  $w\gg \frac{N}{J}$ see  \cite{lunkin2018sachdev,lunkin2020perturbed}. For the opposite case we should limit ourselves  by times $t\ll \frac{N}{J}$ \cite{bagrets2016sachdev}.  

\section{OTOC}
\label{sec: OTOC}
 
The main idea of the paper is to consider the following Ansatz
for  $f(u)$:
\begin{eqnarray}
e^{f(u)}=\frac{e^{u}+a(u)}{b(u)e^u+1}.
\label{eq: ansatz}
\end{eqnarray}
Here  $a(u)$ and $b(u)$ are slow variables i.e. $a^\prime(u)\ll a$ and $b^\prime\ll b$. Using this ansatz we can write an expression for the field $g$ which follows from the expression below:
\begin{eqnarray}
&4\sinh^2\left(\frac{u_1-u_0}{2}\right)\frac{f^{\prime}_{\sigma_1,\textbf{r}}(u_1)f^{\prime}_{\sigma_0,\textbf{r}}(u_0)}{4\sinh^2\left(\frac{f_{\sigma_1,\textbf{r}}(u_1)-f_{\sigma_0,\textbf{r}}(u_0)}{2}\right)}\approx  &\left[1+2\frac{a_{\sigma_1,\textbf{r}}(u_1)-a_{\sigma_0,\textbf{r}}(u_0)}{e^{u_1}-e^{u_0}}+2\frac{b_{\sigma_1,\textbf{r}}(u_1)-b_{\sigma_0,\textbf{r}}(u_0)}{e^{-u_1}-e^{-u_0}}\right]^{-1}
\end{eqnarray}
We have neglected derivatives of the fields $a$ and $b$ as these fields are slow. We also have neglected the terms higher than linear in slow fields as these fields are small ( $a b\propto N^{-1/2}$).  We saved terms with exponentially big coefficients such terms could be of the order of unity.
The following auxiliary integral will be used below:
\begin{eqnarray}
a^{-\gamma}=\int\limits_0^{\infty} s^{\gamma-1} e^{-s a} \frac{ds}{\Gamma(\gamma)}.
\label{eq:usefull integral}
\end{eqnarray}
Using this integral  (\ref{eq:usefull integral})  twice we can write OTOC as:
\begin{eqnarray}
\Tilde{F}(u_1,u_2,u_3,u_4)= \int\limits_0^\infty \frac{ds_o ds_e}{\Gamma(2\Delta)^2} (s_o s_e)^{(2\Delta)-1}e^{-s_e-s_o}\langle e^{i S_j}\rangle \nonumber \\
S_J=\sum\limits_{\textbf{r}}\int du \left( \hat{j}^T_{a,{\textbf r }}(u)\hat{a}_{{\textbf r}}(u)+\hat{j}^T_{b,{\textbf r }}(u)\hat{b}_{{\textbf r}}(u)\right) 
\end{eqnarray}
Here we introduced columns $\hat{a}$ and $\hat{b}$ which contain fields $a$ and  $b$  respectively. These columns have four components accordingly to parts of the double Keldysh contour.  The column  $\hat{j}_a$  has the following form:
\begin{eqnarray}
\hat{j}_a = -i\left(\begin{smallmatrix} \frac{-2s_o\delta(u-u_1)\delta_{{\textbf r },{\textbf r }_1}}{e^{u_1}-e^{u_3}}\\ \frac{-2s_e\delta(u-u_2)\delta_{{\textbf r },{\textbf r }_2}}{e^{u_2}-e^{u_4}} \\ \frac{2s_o\delta(u-u_3)\delta_{{\textbf r },{\textbf r }_1}}{e^{u_1}-e^{u_3}}\\  \frac{2s_e\delta(u-u_4)\delta_{{\textbf r },{\textbf r }_2}}{e^{u_2}-e^{u_4}} \end{smallmatrix}\right).
\end{eqnarray}
The column $\hat{j}_b$ could be written similarly. 

Now we could write an action for slow fields.  As fields   $f$ defined by Eq.  (\ref{eq: ansatz}) are close to the saddle-point solution we can use a quadratic action to find an action for fields $a$ and $b$.  It is convenient to introduce the field $\delta f$ as $f=u+\delta f(u)$. Using this field we can write quadratic action. A similar action was obtained in ref.  \cite{lunkin2021high,lunkin2022non}  and it has the form:

\begin{eqnarray}
S_2 = \frac{1}{2}\int\frac{ d\Omega d^d {\textbf{p}}}{(2\pi)^{d+1}} \delta \hat{f}^\dagger_{{\textbf{p}}}(\Omega) \left[\hat{\mathcal{G}}(\Omega,{\textbf{p}})\right]^{-1} \delta  \hat{f}_{{\textbf{p}}}(\Omega), \nonumber\\ 
\left[\hat{\mathcal{G}}(\Omega,{\textbf p })\right]^{-1} = \left[\hat{\mathcal{G}}^{0}(\Omega)\right]^{-1}- \hat{\Sigma}(\Omega,{\textbf{p}}).
\end{eqnarray}
Here $d$ is a dimension of the lattice we consider.
 $\mathcal{G}$ is a Green's function for soft modes and  $\mathcal{G}^0$ is a Green's function for soft modes in the SYK model. $\Sigma$ is a self-energy originating from the tunneling terms. These propagators have a matrix structure as we have a set of functions $\delta f$  instead of one function. We also use frequency-momentum representation.  As we work in the system in the thermal  equilibrium all components of the Green's functions could be found from the retarded one which has the following form:  
\begin{eqnarray}
\left[\mathcal{G}_R(\Omega,{\textbf{p}})\right]^{-1} = \left[\mathcal{G}_R^{0}(\Omega)\right]^{-1}- \hat{\Sigma}_R(\Omega,{\textbf{p}})  \qquad  \mathcal{G}_R^{0}(\Omega) = \frac{1}{C_J\Omega^2(\Omega^2+1)} , \nonumber \\ 
\Sigma_R(\Omega,{\textbf{p}})= \frac{C_w}{2} \sum\limits_{\delta{\bf r}} \Bigl\{\Omega^2 \psi(\Omega) +\frac{1}{2}\sin^2\left(\frac{{\bf p}\delta{\bf r}}{2} \right)\left(2(1+\Omega^2)+(1+2\Omega^2)\psi(\Omega)\right)\Bigr\}, \nonumber \\
\psi(\Omega)=\Psi\left(\frac{1}{2}-i\Omega\right)-\Psi\left(-\frac{1}{2}\right)\qquad
\Psi(z)=\partial_z\ln\Gamma(z)
\end{eqnarray} 
Action for slow fields $a(\Omega)$ and $b(\Omega)$ defined only for  $\Omega\ll1$ as they are slow. This action has the following  form:
\begin{eqnarray}
S_{2,ab} = \int\frac{ d\Omega d^d {\textbf{p}}}{(2\pi)^{d+1}}  \hat{a}^\dagger_{{\textbf{p}}}(\Omega) \left[\hat{\mathcal{G}}(i+\Omega,{\textbf{p}})\right]^{-1}   \hat{b}_{{\textbf{p}}}(\Omega)
\end{eqnarray}
Using this action we can perform functional integration over these fields:
\begin{eqnarray}
\langle e^{i S_j} \rangle  \approx \exp\{-s_e s_o z\}  
\label{eq:average}
\end{eqnarray}
Where 
\begin{eqnarray}
z\equiv\frac{i}{4}\frac{f_\alpha({\bf r}_1-{\bf r}_2)}{C_J+\sum\limits_{\delta{\bf r}}C_w\left(\frac{\pi^2}{8}-1\right)}\frac{\sinh\left(\frac{u_4+u_2-u_1-u_3}{2}\right)}{\sinh\left(\frac{u_4-u2}{2}\right)\sinh\left(\frac{u_3-u1}{2}\right)},
\label{eq:z}
\end{eqnarray}

\begin{eqnarray}
f_\alpha({\bf r})\equiv \int_{BZ} \frac{d^d {\bf p}}{(2\pi)^d}\frac{e^{i{\bf r}{\bf p}}}{1+\alpha \sum_{\delta{\bf r}} \sin^2\left(\frac{{\bf p}\delta{\bf r}}{2} \right) }, 
\label{eq:f}
\end{eqnarray}

\begin{eqnarray}
\alpha\equiv\frac{\sum\limits_{\delta{\bf r}}C_w\frac{\pi^2}{8}}{C_J+\sum\limits_{\delta{\bf r}}C_w\left(\frac{\pi^2}{8}-1\right)}=  \left[\frac{8\alpha_S}{d\pi^2} \left(\frac{2\pi T}{w}\right)^2 +(1-\frac{8}{\pi^2})\right]^{-1}.
\label{eq:alpha}
\end{eqnarray}
 
The sign  $\approx$ used in (\ref{eq:average}) means that we have neglect terms proportional to $s_e^2$ and  $s_o^2$  due to their "locality". It means that they depend on difference  $u_1,u_3$ or  $u_2,u_4$ and does not depend on $\delta u\equiv u_1-u_2$ as a result they do not contain exponentially large factor.Let us note that, the function $z$ growth exponentially with increasing the difference $\delta u$. On the other hand, the function $f$ decays exponentially for large $|{\textbf r}|$.  

Finally, we can write an explicit formula for the correlator: 
\begin{eqnarray}
&\Tilde{F}(u_1,u_2,u_3,u_4)= \int\limits_0^\infty \frac{ds_o ds_e}{\Gamma(2\Delta)^2} (s_o s_e)^{(2\Delta)}e^{-s_e-s_o} e^{-s_e s_o z}= \nonumber \\
&\int\limits_0^\infty \frac{s^{2\Delta-1} }{\Gamma(2\Delta)} \frac{ e^{-s}}{(1+s z)^{2\Delta}}ds= \frac{U\left(2\Delta,1,\frac{1}{z}\right)}{z^{2\Delta}}=\begin{cases}1-4 z \Delta^2 \quad z\ll1 \\ \frac{\ln(z)}{z^{2\Delta} \Gamma(2\Delta)} \quad z\gg1 \end{cases}. 
\label{eq:answer}
\end{eqnarray}

Here $U$ is a confluent hypergeometric function uniquely determined by its asymptotic behaviour.

This expression together with Eqs. (\ref{eq:z}-\ref{eq:alpha}) are the main results of the presented Letter. Let's review the main properties of this result. 

Firstly, assuming $w=0$ we will obtain the result for OTOC for the SYK model. This result coincides with formula (6.10) of the ref.  \cite{maldacena2016conformal} and also follows from ref.  \cite{gu2022two}. Let us note, that our technique is different from the  methods used in papers \cite{maldacena2016conformal,gu2022two}. We assume that it allows us to use the ansatz    (\ref{eq: ansatz}) for the case $w\ne0$.  

Secondly, the integral    $f_{\alpha}({\bf r})$ determines the dependence of the OTOC on $\delta {\bf r}$. This integral decay exponentially with increasing  $|\delta {\bf r}|$ but this decay depends also on the direction of the vector  $\delta {\bf r}$. For big $\alpha$ integrals come from the region with small $p\ll\frac{1}{a}$. Us a result, correlation function does not depend on the direction of the  $\delta {\bf r}$.   In the general case, the information about perturbation spreads ballistically. It coincides with the results from refs\cite{aleiner2016microscopic,nahum2018operator,von2018operator}.  

Thirdly, formula (\ref{eq:answer}) leads us to the following observation. For  $z\gg1$ the correlator is small $\Tilde{F}\approx 0$ it means that points from this area "know" about applied perturbation whereas points where  $z\ll1$ and  $\Tilde{F}\approx 1$, respectively,  do not know  about perturbation.  Let us discuss it more deeply. 

We fix the direction of the vector  $\delta {\bf r}$  and consider $z$ as a function of the  $|\delta {\bf r}|$. For sufficiently  large times  and big distance we can write the next asymptotic expression  $z\propto e^{\lambda_L (|\delta t|- \frac{\delta {\bf r}}{v})-\ln(N)}$. Here   $v$ is a parameter with dimension of a speed.  Let us consider the area where $z\sim1$, assuming time difference is fixed. This area is quite narrow in comparison with areas where $z\ll 1$ or $z\gg 1$.  This thick border-line region moves with constant speed $v$ with increasing the time difference. 

Let us consider several examples of this motion.

In general case with arbitrary dimension, universal behaviour of  $f$ exist only for region $\frac{r^2}{a^2}\gg \alpha\gg1$. Here $a$ is a lattice constant. In this case  $f\sim\exp\left\{-\sqrt{\frac{2}{\alpha}}\frac{r}{a}\right\}$. It leads to the universal answer for the speed of the front:
\begin{eqnarray}
v=2\pi T \sqrt{\frac{\alpha}{2}}a,
\end{eqnarray} 
This speed also does not depend on the directions.

In the one dimensional system, we can say that ($ {\textbf r}=(n a)$)as a result  the following asymptotic behavior of $f$ could be observed: 
\begin{eqnarray}
f^{(1D)}_{\alpha}({\bf r})=\begin{cases} \left(\frac{\alpha}{2}\right)^{|n|} \quad \alpha\ll1 \\ \frac{1}{\sqrt{2\alpha}}\left(1-\sqrt{\frac{2}{\alpha}}\right)^{|n|}\quad \alpha\gg1 \end{cases}.
\end{eqnarray}

In the two dimensional case with  $\alpha\ll1$:
\begin{eqnarray}
f^{(2D)}_{\alpha\ll1}({\bf r})=\left(\frac{\alpha}{2}\right)^{|n|+|m|} C_{|n|+|m|}^{|m|}
\end{eqnarray}
Here the vector  ${\bf r}=(n a,m a)$, as one can see, this function strongly depend on the direction of this vector  ${\bf r}$. For $\alpha \gg1$  function $f$ has the following form:
\begin{eqnarray}
f^{(2D)}_{\alpha\gg1}({\bf r})= \frac{K_0\left(\sqrt{\frac{2}{\alpha}}\frac{r}{a}\right)}{\pi \alpha}
\end{eqnarray}

Now, we can consider the dependence of the parameter $\alpha$ on the temperature of the system. The parameter $\alpha$ depends on the ratio $\frac{T}{w}$ only  and does not depend on $J$, the largest energy scale in the problem.  For sufficiently large temperatures  $T\gg w$ we can write: $\alpha\sim \left(\frac{w}{T}\right)^2$. As a result we observe strong dependence on the  temperature and the parameter $\alpha$ is small.  In opposite case $w\gg T$ the parameter is almost constant  $\alpha\approx \alpha_0=\frac{\pi^2}{\pi^2-8}$ and universal as it  does not depend on the temperature and other parameters of the system.
Note that  $\alpha_0 \approx5.2\gg1$ 1 so front motion is nearly isotropic at $ T\ll w$.

\section{Conclusion}

In the presented article we have studied the behavior of the OTOC for the system of quantum dots. Dynamics inside dots is described by the SYK Hamiltonian with the typical scale $J$. We also assume the tunneling between dots with typical amplitude $w$.  In this article, we have shown that the Lyapunov exponent of this system reaches the maximal possible value   $\lambda_L=2\pi T$ as in the original SYK model. We also studied the dependence of the OTOC on the spatial parameters. We showed that 
There are two areas 1) area "informed" about perturbation with $(\Tilde{F}\approx 0)$  and 2) area which does know nothing about applied perturbation with $F\approx 1$. There is a border between these two regions which moves ballistically with increasing of time.  The details of moving, in the general case, depend on the parameters of the system, lattice, and temperature.  However, for $w\gg T$ the border is almost spherical   and moves with the constant speed $v_f=2\pi T \sqrt{\frac{\alpha_0}{2}}a$ which depends only on the temperature and the lattice constant.  

It is important to notice that soft modes lead to qualitatively new behavior at moderately low temperatures  $w\gg T\gg T_{FL}\sim\frac{w^2}{J}$; similar results were found in \cite{lunkin2021high,lunkin2022non}.

\section*{ Acknowledgements}

I am grateful to Mikhail Feigel’man and  Alexei Kitaev for discussions during all stages. I am also grateful to Konstantin Tikhonov for the discussion of the properties of the OTOC. 

Research  was partially supported by the Basis Foundation, by the
Basic research program of the HSE and by the RFBR grant № 20-32-90057.

.

\bibliography{syk}

\end{document}